\documentstyle[aps,multicol,epsfig,psfig,prl]{revtex}
\begin{document}

\title{Force-Induced Melting  and Thermal Melting of a 
Double-Stranded Biopolymer}
\author{Haijun Zhou}
\address{Institute of Theoretical Physics, The Chinese Academy of Sciences,
 P. O. Box 2735, Beijing 100080, China}
\date{July 3, 2000}

\maketitle
\widetext
\begin{abstract}
As a prototype of  systems bearing a localization-delocalization 
transition, the strand-separation (melting) process in a  double-stranded
biopolymer is studied by a mapping to a quantum-mechanical problem with
short-ranged potentials. Both the bounded
and the extensive eigenmodes of the corresponding 
Schr\"{o}dinger equation are considered and  exact expressions for 
the configurational partition function and 
free energy are obtained. The force-induced melting
is a first order  phase transition process, while the thermal
melting is found to be second order.  Some scaling exponents governing 
thermal melting are  given.
\end{abstract}

\widetext
\begin{multicols}{2}

DNA melting, the strand-separation of a DNA double-helix,
is an  issue with both practical biological significance 
(because it is closely
related to DNA replication and gene transcription) 
and pure academic interest.
The first realistic  model of DNA melting was proposed
by de Gennes in 1969 \cite{degennes1969} and reintroduced by 
Peyrard and Bishop twenty years later\cite{peyrard1989}.
The essential advance of the de Gennes-Peyrard-Bishop approach, distinguishing
it from earlier efforts based on Ising-like models\cite{poland1966},  is that (i) the 
continuous degrees of freedom for the configurational fluctuation of a DNA and
(ii) the short-ranged hydrogen bonding  between its two complementary
strands have been explicitly incorporated. 
The DNA molecule is considered as consisting of
two flexible Gaussian chains with short-ranged  on-site interactions
between them; its statistical property
can be studied by a mapping to a weakly-bounded quantum mechanical problem
\cite{degennes1969,zhou2000}. 
The corresponding Schr\"{o}dinger equation has a finite number of bounded 
eigenstates and a continuous series of extensive ones, representing
respectively localized and delocalized eigenmodes of structural fluctuations.
In fact, it has been well recognized that problems such  DNA melting, 
flux line depinning in superconducting shells\cite{hatano1996}, 
adsorption of  polymeric 
materials onto a surface\cite{degennes1969,grosberg1986}, and some wetting 
phenomena\cite{kroll1983},  
are all governed by the  competition between enthalpy-favoring 
localized states and  entropy-favoring delocalized ones. 

We study the melting of a double-stranded biopolymer as a prototype of
such many phenomena. Earlier studies have focused mainly on
thermal 
melting\cite{degennes1969,peyrard1989,dauxois1993,zhang97,cule1997}, here both 
the thermal and the force-induced melting
processes are discussed; and different some previous efforts
\cite{peyrard1989,hatano1996,bhatta1999,lubensky2000} 
which considered only  the 
localized ground-state, we explicitly incorporate all the bounded and 
delocalized  eigenmodes of fluctuations (we are grateful to Prof. D. R. Nelson
for informing us that in \cite{hatano1997} a similar treatment has been
performed in studying votex pinning). This makes it possible for us
to obtain exact expressions for the partition function and free energy of the
system.
We rigorously show  that the stretch-induced
melting  is a first order structural phase transition while that induced by
heating is of  second order. The phase diagram for the double-stranded polymer
is  exactly obtained, and some scaling laws governing thermal melting 
are given.  The effects of sequence heterogeneity and base-pair
stacking are 
also briefly discussed.


The model double-stranded biopolymer  is formed by
two Gaussian chains. Each strand contains
$N+1$ beads,  with a harmonic 
attraction between any two consecutive ones;  and between each pair of beads 
of the two strands with the same index there is  a short-ranged
interaction potential $V$. The model energy is
\begin{eqnarray}
 &&{{\cal H}} =\sum\limits_{i=1}^2\sum\limits_{n=1}^N {\kappa \over 2}
({\bf r}_n^{(i)}-{\bf r}_{n-1}^{(i)})^2+\sum\limits_{n=0}^N V({\bf r}_n^{(1)}
-{\bf r}_{n}^{(2)}) \nonumber \\
 && = \sum\limits_{n=1}^N \left[\kappa ({\bf R}_n -{\bf R}_{n-1})^2
+{\kappa\over 4}({\bf r}_n-{\bf r}_{n-1})^2 \right]
+\sum_{n=0}^N V({\bf r}_n). \label{eq:energy1}
\end{eqnarray}
Here, ${\bf r}_{n}^{(i)}$ denotes the position vector of 
the $n$th bead in the $i$th strand; and ${\bf r}_{n}={\bf r}_{n}^{(1)}
-{\bf r}_{n}^{(2)}$ and  ${\bf R}_{n}=({\bf r}_{n}^{(1)}+{\bf r}_{n}^{(2)})/2$ 
are, respectively, the distance between a pair of beads of the two 
strands and the   center-of-mass position of these two beads. 
For computational simplicity, in the following we discuss only the 
one-dimensional case of model Eq.~(\ref{eq:energy1}). 
The  principal  conclusions of this
work is independent of dimensionality as well as  
the particular forms for the 
short-ranged potential $V$, since  the 
underlining physics, the competition between enthalpy in the localized
states and entropy in the extended states, is reserved.
We first discuss the situation of symmetric potentials
and assume the short-ranged attraction to be 
$\delta$-form,  $V(r)=-\gamma \delta(r)$.  
At index $n=0$  the two strands intersect each other, i.e., $R_0=r_0=0$.
The statistical weight for 
the center-of-mass position  at the other end
to be equal to $R$ is  easily known: $Z_{R}(R,  N)=
(\beta \kappa/\pi N)^{1/2} \exp(-\beta \kappa R^2/N)$, where $\beta=
1/k_B T$ with $T$ being the temperature. 

The statistical weight $Z_{r}(r, N)$ for the 
relative distance $r$  between the two strands at index $N$   
is governed by  the 
 Schr\"{o}dinger equation\cite{degennes1969,zhou2000}
\begin{equation}
\label{eq:green1}
{\partial Z_r(r, N) \over \partial N}=\left[{2 \partial^2 \over 
\beta \kappa\partial r^2} -\beta V(r)\right] Z_r(r, N),
\end{equation}
with the initial condition $Z_r(r,0)=\delta(r)$. Equation~(\ref{eq:green1})
corresponds to a weakly-bounded statistical system with only a finite
number of localized eigenstates (in the case of $\delta$-form attraction
used here, this number is unity); the ground
eigenfunction  of Eq.~(\ref{eq:green1}) is  
$\phi_b(r)\propto \exp(-\eta |r|/2)$, where $\eta=\kappa \gamma
\beta^2$. The  statistical weight $Z_r(r,N)$  is expressed as
\begin{eqnarray}
Z_r(r, N)&=&\exp(N \eta^2 /4 \kappa\beta) \phi_b (r)\phi_b(0) \nonumber \\
 & & + \int\limits_{-\infty}^{\infty} d \lambda \exp[-N \epsilon({\lambda})]
\phi(\lambda, r)\phi^*(\lambda,0),
\label{eq:partition1b}
\end{eqnarray}
where $\epsilon(\lambda)$ and $\phi(\lambda, r)$ are the eigenvalue
and the eigenfunction of the extensive eigenstate of 
Eq.~(\ref{eq:green1}) with wave 
number $\lambda$\cite{note1}. 
For the force-fixed ensemble with an external stretching  
$f$  acting on the $N$th bead of the first
strand, the partition function  is 
$\Xi(f,N) =\int d R_N \int d r_N
\exp(\beta f R_N +(1/2) \beta f r_N) Z_R(R_N, N) Z_r(r_N, N)$. 
Since the first term of $Z_r(r, N)$ is proportional to 
$\exp(-\eta |r|/2)$, it seems that  when $f\geq \eta/\beta$ this
integral  will turn to be divergent. 
Earlier  studies\cite{bhatta1999,lubensky2000}  considered only the ground eigenstate and
therefore  took such a divergence as signifying the occurrence 
of force-induced melting process. Actually, however, there is no 
divergence problem. After
taking into account of the second term in Eq.~(\ref{eq:partition1b}), this 
term  is  canceled  out by a term resulted from the integral. 
The correct form of the statistical
weight  is 
\end{multicols}
\widetext
\begin{eqnarray}
Z_r(r, N) &=& \large{(}{\beta \kappa \over 4 \pi N}\large{)}^{1/2}
          \exp(-{\beta\kappa \over 4 N} r^2)+
{\eta \over 4 \pi} \exp({N \eta^2 \over 4 \beta\kappa})\times \nonumber \\
 & &\left\{\int\limits_0^N \sqrt{{\pi\over 4\beta\kappa N^\prime}}
\eta \exp(-{\beta\kappa\over 4 N^\prime} r^2-{N^\prime \over 4\beta \kappa}
\eta^2) d N^\prime 
+\int\limits_0^N \sqrt{{\pi\beta\kappa \over N^\prime}} {|r|\over 2 N^\prime}
\exp(-{\beta\kappa \over 4 N^\prime} r^2 -{N^\prime \over 4\beta \kappa}
\eta^2) d N^\prime \right\};
\label{eq:partition1}
\end{eqnarray}

\widetext
\noindent 
and  the total partition function is thus
%
\begin{equation}
\Xi(f,N)=
\exp({N (\beta f)^2 /4 \beta\kappa})\left[({\eta \over \eta-\beta f} +A_0)
\exp({N\eta^2 \over 4\beta \kappa})+{\beta f \over \beta f-\eta}
\exp({N \beta^2 f^2 \over 4\beta\kappa})\right], 
\label{eq:free1}
\end{equation}
\widetext
\begin{multicols}{2}
\noindent
where $A_0=(2/\sqrt{\pi})\int_0^\infty d t^2\exp(-t^2) 
\int_{\beta f t/\eta}^\infty
d y \exp[- ( y^2 - \beta^2 f^2 t^2/ \eta^2)]$ is an small quantity.
Equation (\ref{eq:free1}) shows that 
in the thermodynamic limit,  the free energy linear 
density is  $g(f)=\lim\limits_{N\rightarrow
\infty} -k_B T \Xi(f,N)/N=  -\kappa \gamma^2 \beta^2/4 
-f^2/4\kappa$ for
$f<f_c$  and $-f^2/ 2 \kappa$ for $f>f_c$, with $f_c=\kappa\gamma \beta$.
Therefore  a first order  phase 
transition occurs at the threshold
force $f_c$.  The inter-beads distance for the first
strand is $f/2\kappa$ for $f<f_c$ and $f/\kappa$ for $f>f_c$ and a 
discontinuity appears at $f_c$. Similarly, the average distance between
the end  beads of the two strands is approximately zero   for 
$f< f_c$ and proportional to $N$ for $f>f_c$ ($\bar{r}_N=N f/\kappa$, 
see Fig.~1).
At $f_c$, $\bar{r}_N=3/\eta+
N\eta /2\kappa\beta$.

The extension-fixed ensemble may be more directly related with actual
experiments\cite{smith1992,cluzel1996,reif1999}: it is much easier for one to  fix the total
extension of the first strand than to fix the  external force. 
The statistical weight for the end-to-end distance of 
the first strand to be fixed at $\sigma N$  is
equal to $\int d R \int d r \delta(\sigma N-R-r/2) Z_R (R, N) Z_r (r, N)$.
$\sigma$ is the inter-beads distance of  the first strand. 
Based on Eq.~(\ref{eq:partition1}), we  
find the free energy density to be  $\tilde{g}(\sigma)=\kappa\sigma^2-
\eta^2/4\kappa \beta^2$ for $\sigma<\sigma_c$, 
$\tilde{g}(\sigma)=\eta \sigma/\beta-\eta^2/2\kappa\beta$ 
for $\sigma_c\leq \sigma\leq 2\sigma_c$,
and $\tilde{g}(\sigma)=\kappa \sigma^2 /2$ for $\sigma> 2\sigma_c$,
with $\sigma_c=\gamma \beta/2$. The 
free energy function is hence 
a piecewise smooth function. Correspondingly,
the average force is  $\bar{f}(\sigma)=2\kappa \sigma$ for
$\sigma<\sigma_c$,   $\bar{f}(\sigma)=\kappa \gamma \beta$ for 
$\sigma_c \leq \sigma\leq 2\sigma_c$, and $\bar{f}(\sigma)=\kappa\sigma$
for $\sigma> 2\sigma_c$ (see Fig.~1). 
The occurrence of a force platform may be striking\cite{cluzel1996,reif1999}.
It indicates that
as the inter-beads distance in the first 
strand reaches $\sigma_c$, melting of
the double-stranded polymer originates from the end point (index $N$) and 
progresses along  the chain until the whole polymer becomes separated.
Stretch-induced melting can be termed as {\it directional 
melting}\cite{bhatta1999,essevaz1997}. The phase diagram of this system is
shown in Fig.~1, it includes a double-stranded 
native region, a single-stranded  denatured region, and a coexisting
region. This  transition is caused by enthalpy-entropy
competition, different from that discussed in 
Ref.~\cite{hemmer1976}, which is caused by the appearance of two
minima in the ground-state eigenfunction.


The above model with  symmetric potential 
does not exhibit thermal melting behavior. 
In the following we improve our model 
by changing the attractive potential in Eq.~(\ref{eq:energy1}) into the
following asymmetric form 
$V(x)=\infty$ for $x\leq 0$ and $V(x)=-\gamma\delta(x-a)$ for
$x> 0$, where $a$  is  a characteristic distance.
For such an asymmetric system,
when the temperature becomes high enough, the localized eigenstate 
disappears. Hence it might be possible
to qualitatively describe the thermal melting of double-stranded
biopolymers.  

We focus on how the distance between the two strands
changes with external stretching or temperature. 
It is convenient for us to assume  that
$r_0=a$.  For this revised model system, we find  that the statistical 
weight for the  relative distance  is
\end{multicols}
\widetext
\begin{equation}
\label{eq:partition2}
Z_r(r, N)=\left\{\begin{array}{ll}
{\zeta \tau \exp(-\zeta \tau/2)\sinh(\zeta\tau r/2 a)
\over a [1-\tau (1-\zeta)]} \exp({N\zeta^2\tau^2\over 4 \kappa \beta a^2})
+\int\limits_0^{\infty} d\lambda {2\lambda^2 \sin\lambda \sin(\lambda r/a)
\exp(-N\lambda^2/\kappa \beta a^2) \over \pi a [\lambda^2-\tau\lambda
\sin 2\lambda+\tau^2 \sin^2 \lambda]}, & \;\; (r<a) \\
{\zeta \tau \exp(-\zeta \tau r/2 a)\sinh(\zeta\tau/2)
\over a [1-\tau (1-\zeta)]}
 \exp({N\zeta^2\tau^2 \over 4 \kappa \beta a^2}) 
+\int\limits_0^{\infty} d\lambda {
2\lambda \sin\lambda [\lambda \sin(\lambda r/a)- \tau \sin\lambda 
\sin\lambda(r/a-1)]
\exp(-N\lambda^2/\kappa \beta a^2) \over \pi a [\lambda^2-\tau\lambda
\sin 2\lambda+\tau^2 \sin^2 \lambda]}. & \;\; (r\geq a)  
\end{array}
\right.
\end{equation}
\widetext
\begin{multicols}{2}
\noindent
In the above expression, $\tau=a \kappa \gamma \beta^2$ and $\zeta$
is the largest solution of 
%
%
$\zeta=1-\exp(-\zeta \tau)$.
%
%
This equation has  a nonzero solution only when $\tau>\tau_c=1$. When
$\tau\leq \tau_c$ the solution is $\zeta=0$. Since
$\tau$ decreases as temperature increases, 
the  polymer's structure
might undergo a transition at temperature $T=T_m=
\sqrt{\kappa \gamma a}/k_B$. 

When  an external force is acting on the  first strand, 
the total partition function is $\Xi(f, N)=
\exp(N\beta^2 f^2/4\kappa \beta)\int_0^{\infty} 
d r \exp(\beta f r/2) Z_r(r, N)$. This integral is always convergent 
for any value of $f$, although the first term of Eq.~(\ref{eq:partition2})
scales as $\exp(-\zeta \tau r/2 a)$. The integrand in
the second term of Eq.~(\ref{eq:partition2}) has two poles at 
$\lambda=\pm \zeta \tau i/2$, hence it will generate a term  
which precisely cancels out the first term in this equation.
To obtain the analytical expression for the partition function $\Xi(f, N)$ 
is nevertheless a demanding task. We evaluate alternatively its Laplace 
transform:
%
\begin{eqnarray}
 & &{\cal L}[\Xi(f,N)](s)=\int\limits_0^\infty d N 
\exp(-s N)\Xi(f, N) \nonumber \\
 & &={8 \sqrt{s (\kappa\beta)^3} a 
[\exp(\beta f a/2)-\exp(-\sqrt{s \kappa \beta} a)]
\over [4 s \kappa\beta -\beta^2 f^2]
[2 \sqrt{\kappa\beta s} a -\tau (1-\exp(-2\sqrt{\kappa\beta s} a))]}.
\label{eq:laplace}
\end{eqnarray}
%
The largest solution of the equation $1/{\cal L}[\Xi(f,N)](s)=0$ 
corresponds to the linear 
free energy density of the polymer system. 
When the temperature is less than $T_m$, the free energy 
density $g(f)= -(\zeta\tau)^2/4\kappa\beta^2 a^2-f^2/4\kappa$ 
for $f<f_c$ and $g(f)=-f^2 /2\kappa$ for $f>f_c$,
where $f_c=\zeta\tau/a\beta$.
Thus, the external force will induce a first order  phase 
transition  at the threshold force $f_c$, which
decreases as the temperature increases (see Fig.~2). This is similar with what we have 
attained with the earlier model. At $T=T_m$, the threshold force decreases
to zero.  The statistical behavior of the extension-fixed
ensemble is also similar with that of the earlier model
and the phase diagram is shown in Fig.~2.

When there is no external force, the free energy density 
is $g=-(\zeta\tau)^2/4\kappa \beta^2 a^2$ for
$T<T_m$ and zero for $T>T_m$. At $T_m$, the free energy and
its first order derivative with temperature is 
continuous but the second order
derivative is not, indicating that the thermal melting at $T_m$ is
a second order continuous phase transition, with a discontinuity in the
specific heat.  The order parameter
for the thermal melting can  chose to be the probability $P_{loc}(n)$ 
for the distance of a pair of beads (with index $n$) 
of the double-stranded polymer to be 
less than the characteristic length $a$.
For the thermal melting process, we can predict based on 
Eq.~(\ref{eq:partition2})  that, as the melting
temperature $T_m$ is approached from below, $P_{loc}\sim 
(T_m-T)^{\tilde{\beta}}$,
with the critical exponent $\tilde{\beta}=1$\cite{binney1992}; 
it is also easy to obtain that
as the temperature approaches $T_m$, the correlation $``$length" 
in $P_{loc}(n)$ between different indices $n$ and $n^\prime$ scales
as $|T_m-T|^{-\tilde{\gamma}}$, with the critical exponent 
$\tilde{\gamma}=2$\cite{binney1992}. 
At $T_m$, $\langle P_{loc}(n) P_{loc}(n^\prime)\rangle_{c} \sim
|n-n^\prime|^{-1+\tilde{\eta}}$\cite{binney1992}, 
but the critical exponent $\tilde{\eta}$ is difficult to
be obtained by the present asymmetric model. Nevertheless, 
we notice that the 
phase diagram for the symmetric (Fig.~1) and the asymmetric model (Fig.~2)
is identical albeit that the symmetric model has $T_m=\infty$. Therefore,
it should be possible for us to obtain a good estimation of $\tilde{\eta}$ 
based on the symmetric model by artificially assuming
$\gamma=\gamma_0 (T_m-T)/T_m$ for $T<T_m$ and $\gamma=0$ otherwise. 
This treatment reduces the melting temperature from infinity 
to $T_m$. For this system  we know from Eq.~(\ref{eq:partition1}) that
$\tilde{\eta}=1/2$,  and we think  it  should be the same  for the
asymmetric model.
 
Can the present approach be extended to consider the 
possible random  variations in the on-site potential $V(r)$ 
(this is caused by the sequence heterogeneity in the case of DNA
\cite{lubensky2000})? 
This is certainly  a challenging problem and beyond the scope of this 
paper.  But we think that  inclusion of such an effect will not alter the 
qualitative behavior of the phase diagram, since in the renormalization
sense, near the transition point the details of the interactions
will be smoothed out\cite{binney1992}.  

It is of interest to ask  whether the inclusion of  base-stacking effect
(by making the parameter 
$\kappa$ in Eq.~(\ref{eq:energy1}) position dependent
as done in Ref.~\cite{dauxois1993}) will change the thermal 
melting from second order
to first order. It seems still be an issue of debate.
An recent work done by Peyrard and coworkers\cite{theodo2000}
answered it  confirmatively, while the numerical work of Cule and Hwa
\cite{cule1997} suggested that the transition behavior is still of second order. 
We noticed that in Ref.~\cite{theodo2000} an force field is first included 
and at the final stage a limiting procedure is performed to make the field
equal to zero. Our present work demonstrates that the property
of the double-stranded system  depends considerably 
on the external field, therefore it
might be helpful for one to carefully evaluate whether the above mentioned
limiting procedure in Ref.~\cite{theodo2000} causes a significant effect.

The author benefits from discussions 
with Xin Zhou and Yong Zhou. He is grateful to Z.-C. Ou-Yang for
encouragement and to  W.-M. Zheng
for bringing Ref.~\cite{hemmer1976} to his attention.

\vskip 0.5cm
\vbox{
\begin{figure}
\begin{center}
\psfig{file=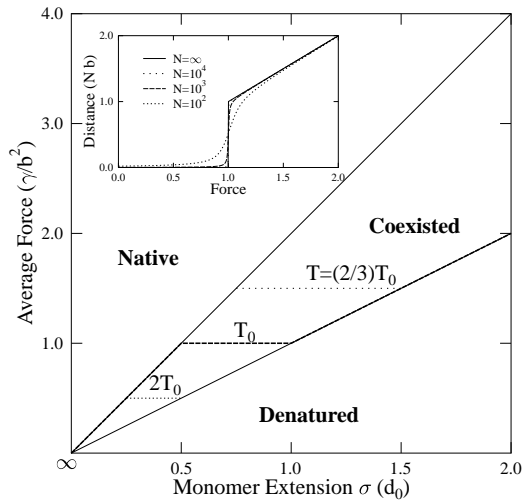, height=3.3in}
\vskip -2.0cm
\narrowtext
\caption{Phase diagram for a double-stranded biopolymer with 
a symmetric potential.  
The broken line shows a force-extension curve at temperature $T_0$. The
phase-coexistence region collapses to a point only when $T\rightarrow 
\infty$, indicating there is no thermal melting. (Inset) The relation
between the  average distance of the ends of the two strands and
the force at temperature $T_0$. 
We set $\kappa=k_B T_0/b^2$ and $\gamma=k_B T_0 d_0$.
\label{fig1} }
\end{center}
\end{figure}
}

\vbox{
\begin{figure}
\begin{center}
\psfig{file=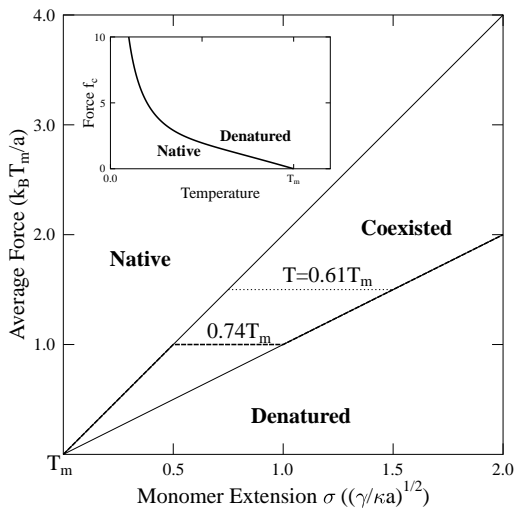, height=3.3in}
\vskip -2.0cm
\narrowtext
\caption{Phase diagram for a double-stranded biopolymer with a
asymmetric potential. This system shows second order thermal
melting behavior at $T_m=\sqrt{\kappa\gamma a}/k_B$. Inset shows 
how the threshold force $f_c$ for melting changes with temperature $T$.
\label{fig2} }
\end{center}
\end{figure}
}

\end{multicols} 

\begin{thebibliography}{99}
\bibitem{note1}
To obtain $\epsilon(\lambda)$ and $\phi(\lambda,r)$, we choose
periodic conditions and use box normalization, the linear dimension of 
which is  taken to be infinity in the last step.
\end{thebibliography}
\end{document}